\documentclass[prd,showpacs,showkeys,nofootinbib,floatfix,
               fleqn,preprint,12pt,tightenlines]{revtex4-1} 

\usepackage{amsmath,amssymb,revsymb,graphicx,dcolumn}

\newcommand{\beq}{\begin{equation}}
\newcommand{\eeq}{\end{equation}}
\newcommand{\beqa}{\begin{eqnarray}}
\newcommand{\eeqa}{\end{eqnarray}}
\newcommand{\bsubeqs}{\begin{subequations}}
\newcommand{\esubeqs}{\end{subequations}}

\begin{document}

\begin{widetext}
\noindent
%
Physics \textbf{2019}, \textit{1}, 321--338  \hfill arXiv:1901.05938
\newline\vspace*{5mm}
\end{widetext}

\title{Vacuum energy decay from a q-bubble}
\vspace*{0mm}
\vspace*{-4mm}

\author{F.R. Klinkhamer}
\email{frans.klinkhamer@kit.edu}
\affiliation{\vspace*{-2mm}Institute for Theoretical Physics,\\
\mbox{Karlsruhe Institute of Technology (KIT),
76128 Karlsruhe, Germany}}

\author{O.P. Santill\'{a}n }
\email{osantil@dm.uba.ar}  
\affiliation{\vspace*{-2mm}Departamento de Matem\'{a}ticas Luis Santal\'{o}n, 
Ciudad Universitaria Pabell\'{o}n I, C1428EGA Buenos Aires, Argentina}

\author{G.E. Volovik}
\email{volovik@ltl.tkk.fi}
\affiliation{\vspace*{-2mm}\mbox{Low Temperature Laboratory, 
Department of Applied Physics,}\\
\mbox{Aalto University, PO Box 15100, FI-00076 Aalto, Finland,}\\
and\\
\mbox{Landau Institute for Theoretical Physics,
Russian Academy of Sciences,}\\
\mbox{Kosygina 2, 119334 Moscow, Russia}\vspace*{0mm}}

\author{A. Zhou}
\email{albert.zhou@kit.edu}
\affiliation{\vspace*{-2mm}Institute for Theoretical Physics,\\
\mbox{Karlsruhe Institute of Technology (KIT),
76128 Karlsruhe, Germany}\vspace*{4mm}}

\begin{abstract}
\vspace*{1mm}\noindent
We consider
a finite-size spherical bubble with a nonequilibrium value of the
$q$-field, where the bubble is immersed in an infinite vacuum
with the constant equilibrium value $q_{0}$ for the $q$-field
(this $q_{0}$ has already cancelled      
an initial cosmological constant).
Numerical results are presented for the time  
evolution of such a $q$-bubble with gravity turned off 
and with gravity turned on.
For small enough bubbles and a $q$-field energy scale 
sufficiently below the gravitational energy scale
$E_\text{Planck}$, the vacuum energy of the
$q$-bubble is found to disperse completely.
For large enough bubbles and a finite value of $E_\text{Planck}$,
the vacuum energy of the $q$-bubble disperses only partially and
there occurs gravitational collapse near the bubble center.
\vspace*{-1mm}
\end{abstract}

\keywords{dark energy,\;
cosmological constant,\;
cosmology
\vspace*{-2mm}}

\maketitle

\section{Introduction}
\label{sec:Intro}

The energy density of the vacuum, the dark energy, and the cosmological
constant are highly debated topics today, as quantum field theory
suggests a typical number that is some 120 orders of magnitude
larger~\cite{Zeldovich1968,Weinberg1988}
than what has been observed~\cite{Tanabashi-etal2018}.
The mismatch is so large and so significant as to make it
the main outstanding problem of modern physics.
However, a similar vacuum energy problem exists in condensed-matter
systems, and its solution may provide a hint for the solution of the
cosmological constant problem. In condensed matter, the zero-point energy
of the quantum fields is fully cancelled by the microscopic (atomic)
degrees of freedom, if the system is in its ground state. If the system
is slightly out of equilibrium, the vacuum energy is not fully
compensated, but its magnitude is determined by the infrared energy scale
rather than by the \mbox{ultraviolet (atomic) energy scale.}

Still, in order to apply
this condensed-matter scenario of the cancellation
of the vacuum energy to the quantum vacuum of our Universe,
we need to know the proper variables to describe this
quantum vacuum.  One example of such a variable is
the four-form field strength used by Hawking
in particular~\cite{Hawking1984}.
The nonlinear extension of this approach, which goes
under the name of $q$-theory~\cite{KV2008a,KV2008b,KV2016-Lambda-cancellation},
demonstrates the nullification of the vacuum energy density
in a full-equilibrium vacuum without matter present.
A small cosmological constant appears if the vacuum is out of equilibrium. Its
value is then determined by infrared physics and is proportional either to
the matter content of the Universe or to the Hubble expansion rate.

While $q$-theory solves the main cosmological problem
(other realizations of the $q$ variable are presented in
Refs.~\cite{KlinkhamerVolovik2016-brane,KlinkhamerVolovik2018-tetrads}
and a one-page review appears as App.~A in   
Ref.~\cite{KlinkhamerVolovik2011-JPCS}), the dynamical
process of equilibration of the vacuum towards the full equilibrium  is
still under investigation. The previously obtained
results~\cite{KV2008b} concern the decay
of an initially homogeneous high-energy state emerging immediately after
the Big Bang. These calculations demonstrated that, with generic
initial conditions, the high-energy state prefers to relax
to a de-Sitter vacuum rather than to the Minkowski vacuum.
On the other hand, the possibility of the final decay of the de-Sitter
vacuum to the Minkowski vacuum is under intensive debate. This is
because of the special symmetry of de-Sitter spacetime; see, e.g.,
Refs.~\cite{Markkanen2018,Matsui2018} and references therein.

One way to circumvent this de-Sitter controversy
is to consider the case that the Big Bang
takes place not over the whole of space but only in a finite
region of space, which is surrounded by equilibrium Minkowski vacuum.
This possibility is also suggested by condensed-matter
experiments~\cite{Ruutu-etal1996},
where a hot spot created within the equilibrium state
finally relaxes to the full equilibrium
by radiating the extra energy away to infinity.

Concretely, we propose to calculate,
in the $q$-theory framework~\cite{KV2008a,KV2008b},
the time evolution of a finite-size spherical
bubble with $q \ne q_{0}$, which is immersed
in an infinite equilibrium vacuum with $q = q_{0}$,
where $q_{0}$ has already cancelled      
an initial cosmological constant $\Lambda$.
The expectation is that the interior field
$q(t,\,r)$ relaxes to $q_{0}$, while the bubble wall
(or its remnant) ultimately moves outwards.
Yet, gravity may hold surprises in store.
Remark that our proposed calculation 
corresponds to the scenario discussed in the
second paragraph of Sec.~V A in Ref.~\cite{KV2008b},
which mentioned the possibility that
``the starting nonequilibrium state could, in turn, be obtained
by a large perturbation of an initial equilibrium vacuum.''
We emphasize that the calculation of the present article  
is the first-ever calculation of the inhomogeneous dynamics 
of the quantum vacuum in the $q$-theory framework.

Before we
start with this calculation, we have three
clarifying comments. The first comment is that
it may be instructive to compare our $q$-bubble to the
vacuum bubble as discussed by Coleman and
collaborators~\cite{Coleman1977,CallanColeman1977,ColemanDeLuccia1980}.
That discussion starts from a classical field theory of a 
fundamental scalar field $\phi(x)$ with nonderivative interactions.
The interactions are, in fact, determined by a
potential term $V(\phi)$ in the action.
The potential $V(\phi)$ is assumed to have various local minima:
one or more ``false'' vacua $\phi_{+,n}$
and the single ``true'' vacuum $\phi_{-}$,
where the ``false'' vacua have a larger energy density $V(\phi_{+,n})$
than the value $V(\phi_{-})$ of the ``true'' vacuum.
Coleman's vacuum bubble, then, corresponds to
a finite-size spherical bubble with ``true'' vacuum inside
and ``false'' vacuum outside (in other words, the energy density
inside is lower than outside). The dynamic behavior of
a single vacuum bubble is that the bubble expands
(cf. Fig.~4 in Ref.~\cite{Coleman1977}) with the true-vacuum
region increasing but, at a given finite time,
the far-away region remaining in a false-vacuum state.
Such a vacuum bubble is essentially different from our $q$-bubble
which has an infinite equilibrium vacuum with $q = q_{0}$ outside
(in Coleman's terminology, ``true'' $q$-vacuum outside).
In a way, the $q$-bubble resembles Coleman's vacuum bubble
with interior and exterior regions switched.
It is clear that, already energetically, the dynamic
behavior of the $q$-bubble will be different from that of
Coleman's vacuum bubble.

The second comment
concerns the different role of a fundamental scalar field $\phi(x)$
and the vacuum variable $q(x)$ for the cosmological constant problem.
In the fundamental-scalar-field approach,
the nullification of the energy density
$\epsilon(\phi)$  
in the equilibrium vacuum
requires fine-tuning~\cite{Weinberg1988}.
In the $q$-field approach, the vacuum is a self-sustained system, which,
in equilibrium, automatically acquires a zero value for 
the thermodynamic  
potential $\widetilde{\epsilon}(q) =\epsilon(q)- q\,d\epsilon(q)/dq$ 
that
enters the Einstein equation by a cosmological-constant-type term.
See, in particular,
the discussion of Sec.~2 in Ref.~\cite{KlinkhamerVolovik2016-brane}.

The third comment expands on the second and      
concerns the actual dynamics of the $q$-field.
At first glance, the dynamical equations used in the present article
are identical to the equations of general relativity coupled
to a ``scalar'' field $q(x)$ with a potential $\rho_{V}(q)$ 
to be defined later.  
The dynamics of a fundamental scalar field $\phi(x)$
interacting with gravity has been extensively studied, 
in particular by Choptuik and
collaborators~\cite{Choptuik1993,MarsaChoptuik1996,HondaChoptuik2002}
(see also Refs.~\cite{Choptuik-etal2015,Cardoso-etal2015}
for two recent reviews on numerical relativity).
In general, however, the $q$-field has only locally the property 
of a scalar field, while it globally obeys a conservation law.
It is precisely this conservation law that makes the four-form field strength
appropriate for the description of the phenomenology of the quantum vacuum. 
All this makes the dynamics of the quantum vacuum essentially different 
from the dynamics of a fundamental scalar field.
This issue will be discussed further in Sec.~\ref{sec:Theory and setup}.

We, now, turn to the calculation of the time evolution of a $q$-bubble.
After a brief review of the theory, we, first, consider a $q$-bubble
with gravity effects turned off and, then, with gravity effects turned on.
Throughout, we use natural units with $c=\hbar=1$
and take the metric signature $(-+++)$.

\section{Theory and setup}
\label{sec:Theory and setup}

In this article, we use $q$-theory
in the four-form-field-strength realization with explicit
derivative terms of the $q$-field in the 
action~\cite{KV2016-q-ball,KV2016-q-DM,KV2016-more-on-q-DM,KlinkhamerMistele2017}.
Specifically, we take the simplest possible
theory with the following action~\cite{KV2016-q-DM}:   
\begin{subequations}\label{eq:action-sigmaLambda-definition-Fq-definition}
 \begin{eqnarray}
S&=&- \int_{\mathbb{R}^4}
\,d^4x\, \sqrt{-g}\,\left(
\frac{R}{16\pi G}   
+\epsilon(q) \right.
\left.
+\frac{1}{2}\,C(q)\, 
g^{\alpha\beta}\, (\nabla_\alpha\, q)\,(\nabla_\beta\, q)
\right),
\label{eq:action}
\\[2mm]
\label{eq:sigmaLambda-definition}
\epsilon(q) &=&  \sigma(q) + \Lambda\,,
\quad
\frac{d \sigma(q)}{d q}\ne 0\,,   
\\[2mm]
\label{eq:Fq-definition}
F_{\alpha\beta\gamma\delta} &\equiv&
\nabla_{[\alpha}A_{\beta\gamma\delta]}\,,
\quad
F_{\alpha\beta\gamma\delta} =
q\,\sqrt{-g} \,\epsilon_{\alpha\beta\gamma\delta}\,,
\end{eqnarray}
\end{subequations}
where $g$ is the determinant of the metric $g_{\alpha\beta}$, 
$R$ the Ricci curvature scalar, 
$G$ a gravitational coupling constant, and
$A$ a three-form gauge field with 
corresponding four-form field strength $F \propto q$
(see Refs.~\cite{KV2008a,KV2008b} and further references therein).
In \eqref{eq:action},  
$C(q) >0$ and $\sigma(q)$ are generic even functions of $q$.
We use the same conventions for the curvature tensors
as in Ref.~\cite{Weinberg1972}.
For the moment, we have omitted in the integrand on the right-hand side
of \eqref{eq:action} the Lagrange density of the fields
of the Standard Model of elementary particle physics.

The Hamilton principle for     
variations $\delta A_{\alpha\beta\gamma}$
and $\delta g_{\mu\nu}$ of
the action \eqref{eq:action-sigmaLambda-definition-Fq-definition}
produces two field equations, a generalized Maxwell equation
involving $d\epsilon(q)/dq$ and the Einstein equation involving
a particular combination of energy-density terms,
\beq\label{eq:epsilon-tilde}
\widetilde{\epsilon}(q) \equiv  
\epsilon(q)- q\,\frac{d \epsilon(q)}{d q}\,.
\eeq
These Maxwell-type and Einstein field equations 
are given by (3) and (5), respectively,
in Ref.~\cite{KV2016-q-DM}.
One particular solution has the flat-spacetime Minkowski
metric,
\begin{subequations}\label{eq:Minkowskimetric-q0-epsilontilde0}
\beq
g_{\mu\nu}(x)=\eta_{\mu\nu}\,,
\eeq
and the constant nonvanishing $q$-field
\beq
q(x)=q_{0}>0\,,
\eeq
where the equilibrium value $q_{0}$ gives 
\beq\label{eq:epsilontilde0}
\widetilde{\epsilon}(q_{0})=\rho_{V}(q_{0})=-P_{V}(q_{0})=0\,,
\eeq
\end{subequations}
with $\rho_{V}$  and $P_{V}$, respectively,
the vacuum energy density and vacuum pressure
entering the Einstein equation (see below). 
Note that the mass dimension of $q_{0}$ is 2
in the four-form-field-strength realization.

The solution of the generalized Maxwell equation     
introduces an integration constant $\mu$ and 
is given by (4) in Ref.~\cite{KV2016-q-DM}.
A particular value for $\mu$ is $\mu_{0}$, which corresponds to
the constant equilibrium value $q_{0}$ of the $q$-field 
[obtained from the condition $\widetilde{\epsilon}(q_{0})=0$] 
and is explicitly defined by
\beq\label{eq:mu0}
\mu_{0} \equiv \left. \frac{d \epsilon(q)}{d q}\,\right|_{q=q_{0}}\,.
\eeq
The solution of the generalized Maxwell equation
now takes the form of a nonlinear Klein--Gordon
equation for the special case of constant $C(q)$,
\beq\label{eq:C-Ansatz}
C(q)  = (q_{0})^{-1}\,.
\eeq
This nonlinear Klein--Gordon equation then reads~\cite{KV2016-q-DM}
\beq\label{eq:nonlin-Klein-Gordon-eq}
(q_{0})^{-1} \, \Box \, q  = \frac{d \rho_{V}(q)}{d q}\,,
\eeq
in terms of the vacuum energy density $\rho_{V}(q)$
defined by
\beq\label{eq:rhoV}
\rho_{V}(q) \equiv \epsilon(q) - \mu_{0}\,q\,,
\eeq
with the constant $\mu_{0}$ from \eqref{eq:mu0}.
Precisely this vacuum energy density $\rho_{V}(q)$ enters
the Einstein equation~\cite{KV2016-q-DM},
\begin{subequations}\label{eq:Einstein-eq}
\beqa
R_{\alpha\beta} - \frac12\, g_{\alpha\beta}\,R &=& 
- 8\pi G\,  
T_{\alpha\beta}^{\,(q)} \,,
\\[2mm]
T_{\alpha\beta}^{\,(q)}
  &=&- \left( \rho_{V}(q) 
+ \frac12\,(q_{0})^{-1}\,\nabla_\gamma \,q \,\nabla^\gamma q   
\right)\, 
        g_{\alpha\beta}
 +(q_{0})^{-1}\, \nabla_\alpha \, q \, \nabla_\beta \, q \,,
\eeqa
\end{subequations}
where $R_{\alpha\beta}$ is the Ricci curvature tensor and 
$T_{\alpha\beta}^{\,(q)}$ the energy-momentum tensor of the $q$-field.

As mentioned in Sec.~\ref{sec:Intro}    
and in Ref.~\cite{KV2016-q-DM}, 
the final dynamic equations \eqref{eq:nonlin-Klein-Gordon-eq}
and \eqref{eq:Einstein-eq} are identical to those of a
gravitating fundamental scalar field $\phi(x)$
with a potential $\rho_{V}(\phi)$
from \eqref{eq:rhoV} with $q$ replaced by $\phi$.
But the constant $\mu_{0}$ entering our two dynamic equations
via $\rho_{V}$
arises as an integration constant for the solution of
an underlying dynamic equation, namely, 
the generalized Maxwell equation obtained by variation of
the three-form gauge field $A$ in the action.
Concretely, the equilibrium value $q_0$ is found to depend on
the cosmological constant $\Lambda$ from
\eqref{eq:sigmaLambda-definition}, 
\bsubeqs\label{eq:q0-Lambda}
\beq\label{eq:q0-Lambda-mu0-Lambda}
q_{0} = q_{0}(\Lambda)\,,
\eeq
and the same holds for
the integration constant $\mu_{0}$ from   
\eqref{eq:mu0},
\beq\label{eq:mu0-Lambda}
\mu_{0} = \mu_{0}(\Lambda)\,.
\eeq
\esubeqs
This point will be clarified by an 
example in the penultimate paragraph of this section.

Remark also     
that the nonlinear Klein--Gordon equation 
\eqref{eq:nonlin-Klein-Gordon-eq} only appears for the
special case of constant $C(q)$ and
constant $G(q)$ 
[here, we have taken $G(q)=G=\text{constant}$].   
The advantage of considering this simplified case
of $q$-theory is that, if necessary, we may appeal to 
established numerical methods~\cite{Choptuik1993,MarsaChoptuik1996,%
HondaChoptuik2002,Choptuik-etal2015,Cardoso-etal2015}
for a gravitating fundamental scalar field $\phi(x)$.
But, here, we will only perform an exploratory numerical
analysis, leaving refinements to the future.

Using $q_{0}$, we introduce the dimensionless
coordinates $(\tau,\,\rho)$ for $(t,\,r)$,
the dimensionless function $f(\tau,\,\rho)$ for $q(t,\,r)$,
the dimensionless constant $u_{0}$ for $\mu_{0}$,
the dimensionless cosmological constant $\lambda$
for the cosmological constant $\Lambda$,
and the dimensionless vacuum energy density $r_{V}(f)$ for $\rho_{V}(q)$.
By abuse of notation, we also have the dimensionless vacuum energy density 
$\epsilon(f)$ for the dimensional quantity $\epsilon(q)$.
Recall that $\mu_{0}$ is the equilibrium value of the
``chemical potential'' $\mu(q)\equiv d \epsilon(q)/d q$ 
corresponding to the
conserved vacuum variable $q$; see Ref.~\cite{KV2008a} for further
discussion.

In order to be specific, we      
take the following \textit{Ansatz} for the dimensionless energy density
$\epsilon(f)$ appearing in the action \eqref{eq:action}:
\beq
\label{eq:epsilon-Ansatz}
\epsilon(f)= \frac{1}{2}\,f^{2}\,\left(\frac{1}{3} \,f^2-1\right)  +\lambda\,,
\eeq
with a dimensionless bare cosmological constant $\lambda\geq 0$
(the case of an arbitrary-sign initial cosmological constant 
$\lambda$ has been  considered in Ref.~\cite{KV2016-Lambda-cancellation}).
The equilibrium condition,
\beq\label{eq:epsilon-tilde-f-is-zero}
\widetilde{\epsilon}(f) \equiv  
\epsilon(f)- f\,\frac{d \epsilon(f)}{d f}=0\,,
\eeq
gives the following constant equilibrium value $f_{0}$ of the $f$-field 
($f_{0}$ is taken to be positive) 
and corresponding ``chemical potential'' $u_{0}$: 
\bsubeqs\label{eq:f0-u0}
\beqa\label{eq:f0}
\hspace*{-2mm}
f_{0}&=&  {\sqrt{\left(1 + {\sqrt{1 + 8\, \lambda}}\right)/2}}
=1 + \lambda + \text{O}(\lambda^2)\,,  
\\[2mm]\label{eq:u0}
\hspace*{-2mm}
u_{0} &\equiv& 
 \left. \frac{d \epsilon(f)}{d f}\,\right|_{f=f_{0}}
\nonumber\\[1mm]
&=&
\frac{1}{3\, {\sqrt{2}}}\,\left( -2 + {\sqrt{
          1 + 8\, \lambda}} \right) \,
           {\sqrt{1 + {\sqrt{1 + 8\, \lambda}}}}
=- \frac{1}{3}   + \lambda + \text{O}(\lambda^2)\,. 
\eeqa
\esubeqs
The dimensionless gravitating vacuum energy density $r_{V}(f)$
corresponding to \eqref{eq:rhoV} is given by
\beqa\label{eq:rV}
r_{V}(f)&\equiv&\epsilon(f)-u_{0}\,f\,,
\eeqa
where the numerical value for $u_{0}$ from \eqref{eq:u0}
holds for the specific  \textit{Ansatz} \eqref{eq:epsilon-Ansatz}.
At equilibrium, the function $r_{V}(f)$ has 
\bsubeqs
\beqa
r_{V}(f_{0})&=&0\,,
\\[2mm]
\left[\frac{d r_{V}(f)}{d f}\,\right]_{f=f_{0}}&=&0\,,
\\[2mm]
\label{eq:inverse-chi0}
\left[f^{2}\,\frac{d^2 r_{V}(f)}{d f\,d f}\,\right]_{f=f_{0}}
&\equiv& 
(\chi_{0})^{-1} 
=
\frac{1 + 8\, \lambda + {\sqrt{1 + 8\, \lambda}}}{2}
=1 + 6\, \lambda + \text{O}(\lambda^2)\,,  
\eeqa
\esubeqs
where $\chi_{0}$ in \eqref{eq:inverse-chi0}
is the dimensionless version of the
equilibrium vacuum compressibility~\cite{KV2008a}.

Observe that $r_{V}(f)$ as defined by \eqref{eq:rV}     
has a direct $\lambda$ dependence 
from the energy density \eqref{eq:epsilon-Ansatz}
and an indirect $\lambda$ dependence 
from the equilibrium value \eqref{eq:mu0}
of the chemical potential, explicitly given by \eqref{eq:u0}.
Let us briefly discuss the implications of this
indirect $\lambda$ dependence.
Write the energy density from \eqref{eq:epsilon-Ansatz}
as 
\bsubeqs
\beq
\epsilon(f;\,\lambda)=s(f)+\lambda
\eeq
and the gravitating energy density from \eqref{eq:rV} as 
\beq
r_{V}(f;\,\lambda)=s(f)+\lambda-u_{0}(\lambda)\,f\,.
\eeq
\esubeqs
Now, consider 
\beq\label{eq:lambdahat}
\widehat{\lambda}=\lambda_{1} + \lambda_{2}\,,
\eeq
for generic positive $\lambda_{1}$ and $\lambda_{2}$.
It then follows that
\bsubeqs
\beq\label{eq:epsilon-lambdahat}
\epsilon(f;\,\widehat{\lambda})
=\epsilon(f;\,\lambda_{1})+\lambda_{2} \,.
\eeq
But the $r_{V}$ behavior is different,
\beq\label{eq:rV-lambdahat}
r_{V}(f;\,\widehat{\lambda})
\ne
r_{V}(f;\,\lambda_{1})+\lambda_{2}\,,
\eeq
\esubeqs
simply because of the shift of the $q$-field equilibrium value $q_0$
if $\lambda_{1}$ is changed to $\widehat{\lambda}$
and the corresponding shift of the
equilibrium chemical potential \eqref{eq:mu0},
as shown by the explicit dimensionless expression \eqref{eq:f0}. 
The additive behavior \eqref{eq:epsilon-lambdahat} 
is what is expected for a fundamental scalar field, but the
behavior \eqref{eq:rV-lambdahat} 
from the composite scalar field $q$ is different.
Precisely this nontrivial behavior of $\rho_{V}(q)$, 
different from the behavior of $\epsilon(\phi)$ 
for a fundamental scalar $\phi$,
allows for the natural compensation of an 
initial  cosmological constant $\Lambda$,
as mentioned in the second comment of Sec.~\ref{sec:Intro}.

\begin{figure}[t]
\vspace*{-0mm}
\begin{center}   
\includegraphics[width=1\textwidth]{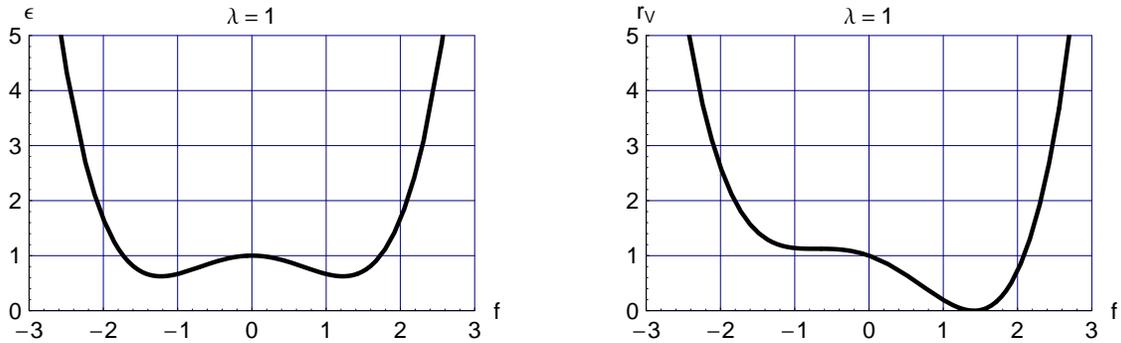}
\end{center}
\vspace*{-5mm}
\caption{Vacuum energy densities $\epsilon(f)$ [on the left] 
and $r_{V}(f)$ [on the right] 
for nonzero cosmological constant $\lambda=1$.
The relevant expression for these  vacuum energy densities are given by
\eqref{eq:epsilon-Ansatz} and \eqref{eq:rV}, with the constant \eqref{eq:u0}.
The vacuum energy density $r_{V}(f)$ is the quantity that gravitates.
}
\label{fig:epsilon-rV-for-lambda-is-1}
\vspace*{-0mm}
\end{figure}

Our numerical calculations will be performed for the case
$\lambda=1$ with $f_{0}=\sqrt{2}$ and $u_{0}=\sqrt{2}/3$ from
\eqref{eq:f0} and \eqref{eq:u0}, respectively. 
The two vacuum energy densities 
are shown in Fig.~\ref{fig:epsilon-rV-for-lambda-is-1}.


\section{Bubble without gravity}
\label{sec:Bubble-without-gravity}

\subsection{Preliminaries}
\label{subsec:Preliminaries}

It is relatively easy to get a result for a special case. 
First, we set 
\beq\label{eq:G-zero}
G=0\,,
\eeq
so that we just have Minkowski spacetime to consider.

Second, we recall from Sec.~\ref{sec:Theory and setup}
that the generalized Maxwell equation~\cite{KV2008a,KV2008b} gives rise to
the nonlinear-Klein--Gordon equation \eqref{eq:nonlin-Klein-Gordon-eq},
which reads explicitly
\beq\label{eq:nonlin-KG-eq}
\Box \, q = q_{0}\,\frac{d \rho_{V}}{d q}\,,
\eeq
with the flat-spacetime d'Alembertian
$\Box\equiv  \eta^{\alpha\beta}\,\partial_{\alpha} \partial_{\beta}
     = -\partial_{t}^{2}+\nabla^{2}$.

Third, introducing spherical coordinates,
the $q$-field of a spherical bubble is given by
\beq\label{eq:spher-symm-q-field}
q=q(t,\,r)\,.
\eeq

Fourth,
we start from a bubble with essentially 
$q_\text{inside}= \widehat{q}\ne q_{0}$
and $q_\text{outside}=q_{0}$.
Outside the bubble, the $q$-field has already compensated the
initial cosmological constant $\Lambda \geq 0$
(the case of an arbitrary-sign initial cosmological constant 
$\Lambda$ has been     
considered in Ref.~\cite{KV2016-Lambda-cancellation}). 
The question, now, is how the inside $q$-field evolves with time.

\subsection{Numerics}
\label{subsec:Numerics-without-grav}

The numerical solution will be obtained by use of the
dimensionless variables introduced in
Sec.~\ref{sec:Theory and setup}.
The partial differential equation (PDE)
from \eqref{eq:nonlin-KG-eq} for the spherically symmetric
$q$-field \eqref{eq:spher-symm-q-field} then reads
\beq\label{eq:flat-spacetime-reduced-nonlinear-KG-eq}
\partial_{\tau}^{2} \, f(\tau,\,\rho)
-\frac{1}{\rho^{2}}\,\partial_{\rho}
\left[\rho^{2}\,\partial_{\rho}\,f(\tau,\,\rho)\right]
=
-\left[ \frac{d}{d f}\,r_{V}(f) \right]_{f=f(\tau,\,\rho)}\,,
\eeq
where $r_{V}(f)$ is given by \eqref{eq:rV} with
\eqref{eq:epsilon-Ansatz} and \eqref{eq:u0}. 
The initial values at $\tau=0$ and   
the boundary conditions at $\rho=0$ and $\rho=\infty$ are
\bsubeqs\label{eq:f-bcs}
\beqa
\label{eq:bcs-f-start}
f(0,\,\rho)&=& f_\text{start}(\rho)\,,
\\[2mm]
\label{eq:bcs-partial-tau-f-start}
\partial_{\tau} f(0,\,\rho) &=& 0\,,
\\[2mm]
\label{eq:f-bcs-rho-origin}
\partial_{\rho}\, f(\tau,\,0)&=&0\,,
\\[2mm]
\label{eq:f-bcs-rho-infinity}
f(\tau,\,\infty)&=&f_{0}\,.
\eeqa
\esubeqs
Practically, we restrict the $\rho$ range
to $\{\rho_\text{min},\rho_\text{max}\}$ with
$\rho_\text{min} \geq 0$ and $\rho_\text{max}<\infty$.
Also, we use the following explicit start function:
\bsubeqs\label{eq:f-start}     
\beqa
\hspace*{-6mm}
f_\text{start}(\rho)&=&
\left\{
\begin{array}{cl}
\widehat{f}\,, &\;\;\text{for}\;\;\rho\in (0,\,\overline{\rho}-1/2)  \,,
\\[1mm]
\widehat{f} + \sin^{4}\Big[\big(\rho - \overline{\rho}+1/2\big)\pi/2\Big]\,
\big(f_{0} - \widehat{f}\big)\,,
&\;\;\text{for}\;\;\rho \in [\overline{\rho}-1/2,\,\overline{\rho}+1/2] \,,
\\[1mm]
f_{0}\,, &\;\;\text{for}\;\;\rho\in (\overline{\rho}+1/2,\infty)\,,
\end{array}
\right.
\eeqa
where, for now, we set $\overline{\rho}=1$ and take
\beqa
\label{eq:f-hat}
\hspace*{-6mm}
\widehat{f} &=& 0\,.
\eeqa
\esubeqs
Note that  
the fourth power of the sine-function in \eqref{eq:f-start}
makes for a continuous second-order derivative
at $\rho=\overline{\rho} \pm 1/2$.


%
\begin{figure}[t]   
\vspace*{-1mm}      
\begin{center}   
\includegraphics[width=1\textwidth]{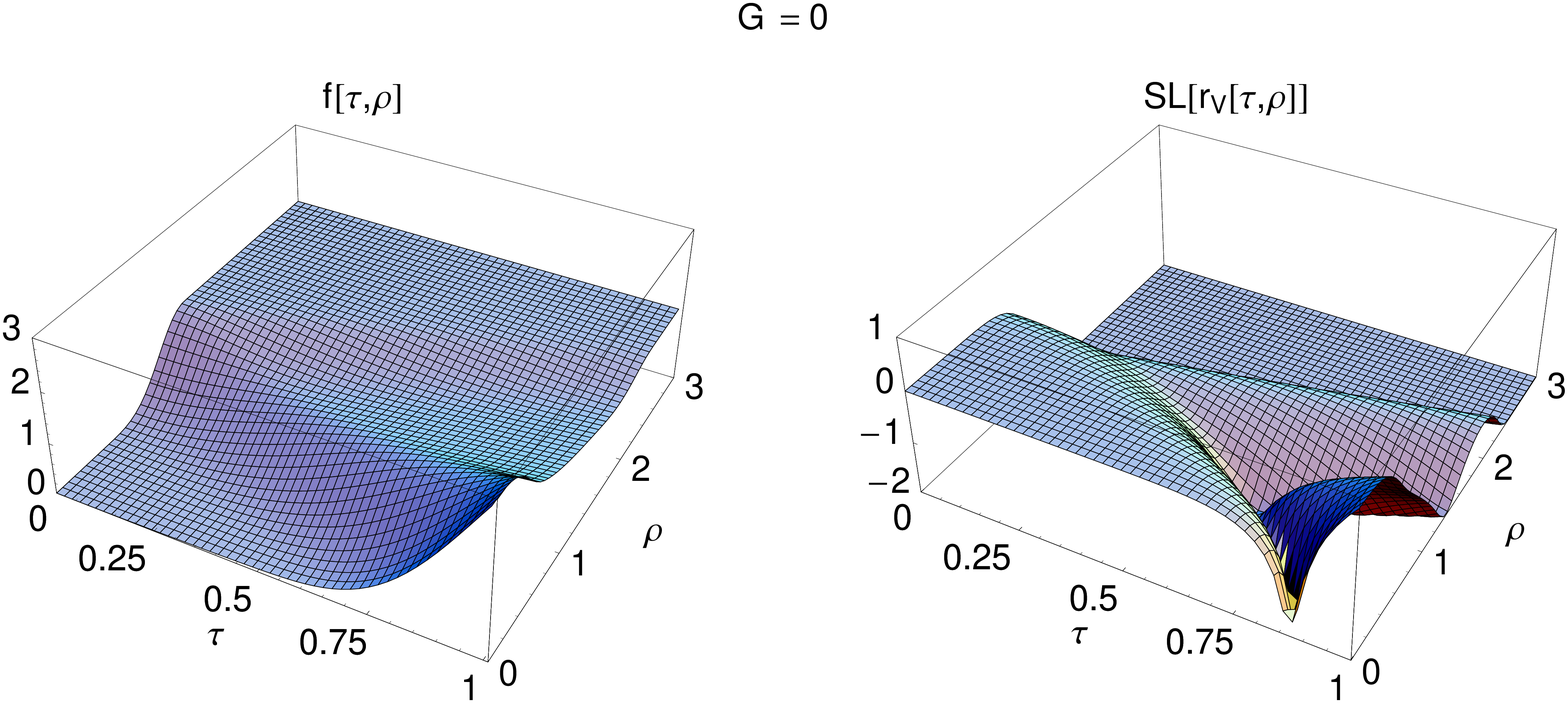}
\end{center}
\vspace*{-5mm}
\caption{Numerical solution of the flat-spacetime
PDE \eqref{eq:flat-spacetime-reduced-nonlinear-KG-eq}
for the case of a nonzero cosmological constant, $\lambda=1$.
The initial values are \eqref{eq:bcs-f-start} from the start function
\eqref{eq:f-start} with $\overline{\rho}=1$
and \eqref{eq:bcs-partial-tau-f-start}.
The boundary conditions are
\eqref{eq:f-bcs-rho-origin} at $\rho_\text{min}=0$
and \eqref{eq:f-bcs-rho-infinity} at $\rho_\text{max}=4$.
The $f(\tau,\,\rho)$ field is calculated over a relatively short time interval,
$\tau \in [0,\,1]$.
Also plotted is the corresponding energy density $r_{V}$,
using the shift-log function defined by
$\text{SL}(x)\equiv\log_{10}(x+0.01)\in [-2,\,\infty)$ for $x\geq 0$.
This vacuum energy density $r_{V}[f]$ for $\lambda=1$
is given by \eqref{eq:rV} with
\eqref{eq:epsilon-Ansatz} and \eqref{eq:u0}.
The vacuum energy density $r_{V}[f]$
is, in fact, the quantity that would gravitate 
if $G$ were nonzero. 
}
\label{fig:two-panel-f-rV-surfaceplots-start-bubble-for-g-0}
\vspace*{2cm}
\vspace*{-0cm}
\begin{center}   
\includegraphics[width=1\textwidth]{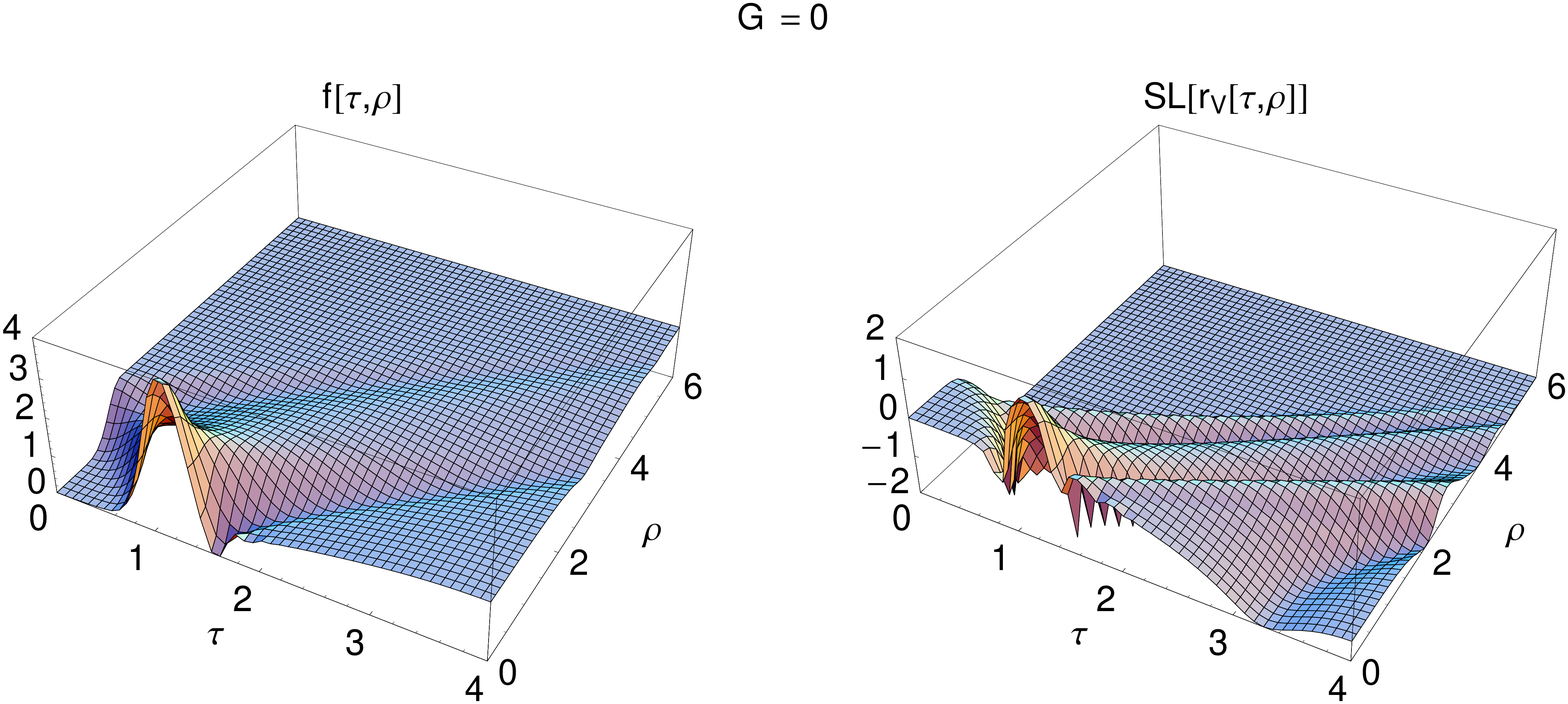}
\end{center}
\vspace*{-5mm}
\caption{Same as Fig.~\ref{fig:two-panel-f-rV-surfaceplots-start-bubble-for-g-0},
but now evolved over a larger time interval,
$\tau \in [0,\,4]$.
}
\label{fig:two-panel-f-rV-surfaceplots-bubble-for-g-0}
\vspace*{0cm}
\end{figure}
\begin{figure}[t]
\vspace*{-0cm}
\vspace*{-0cm}
\begin{center}   
\includegraphics[width=1\textwidth]{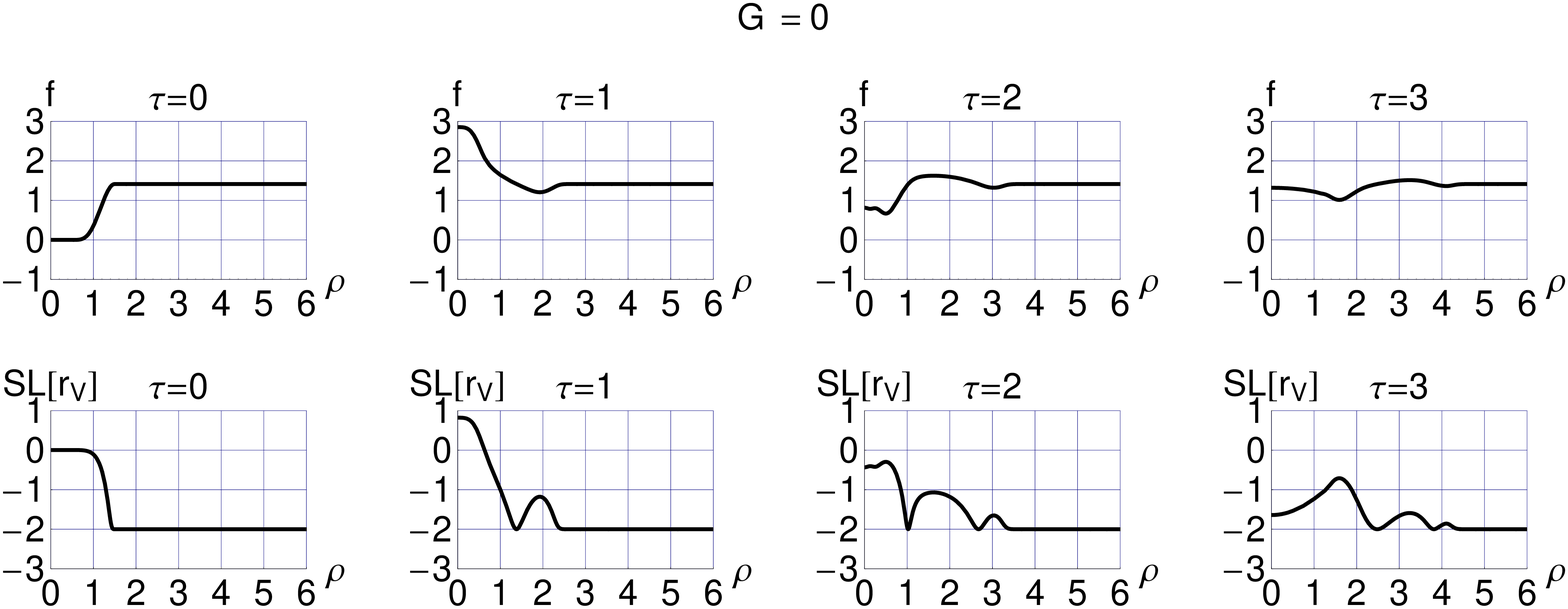}
\end{center}
\vspace*{-6mm}
\caption{Four  time-slices from the numerical solution of
Fig.~\ref{fig:two-panel-f-rV-surfaceplots-bubble-for-g-0}.}
\label{fig:two-panel-f-rV-timeslices-bubble-for-g-0}
\vspace*{-1mm}
\end{figure}

The general behavior of the numerical solution is displayed 
in Figs.~\ref{fig:two-panel-f-rV-surfaceplots-start-bubble-for-g-0}
and \ref{fig:two-panel-f-rV-surfaceplots-bubble-for-g-0}
and four time-slices are given
in Fig.~\ref{fig:two-panel-f-rV-timeslices-bubble-for-g-0}.
These results show the disappearance
of the bubble ``domain-wall'' and the start of the outward motion
of its remnant.
Observe, in Fig.~\ref{fig:two-panel-f-rV-surfaceplots-bubble-for-g-0},
both spatial $r_{V}$ oscillations (for example, at $\tau=4$)
and temporal $r_{V}$ oscillations (for example, at $\rho=0$).
The temporal $r_{V}$ oscillations were first observed for a
homogeneous context in Ref.~\cite{KV2008b}, but new here is that
energy can escape towards the surrounding unperturbed space.

These numerical results demonstrate that the out-moving $r_{V}$ disturbance has a
rapidly decreasing amplitude.
Incidentally, the quality of the numerical solution can be
monitored by evaluating the numerical value of the
integral of motion (energy) corresponding to
the field equation \eqref{eq:flat-spacetime-reduced-nonlinear-KG-eq};
see also Sec.~\ref{subsec:Dimensionless-PDEs}.
\vspace*{-2mm}

\subsection{Discussion}
\label{subsec:Discussion-without-grav}
\vspace*{-1mm}

The numerical results of
Sec.~\ref{subsec:Numerics-without-grav}
show two characteristics of the $q$-bubble time evolution
in the absence of gravitational effects:
\vspace*{-1.5mm}
\begin{enumerate}
\item initially, the bubble wall gives rise to
both out-moving and in-moving disturbances of the dimensionless
vacuum energy density $r_{V}$, where the in-moving disturbance
makes for an increased energy density at the center;
\vspace*{-1.5mm}
\item
ultimately, there is an out-moving $r_{V}$ disturbance with a rapidly
diminishing amplitude (asymptotically, $r_{V} \sim 1/\rho^2  \sim 1/\tau^2$
from energy conservation).
\end{enumerate}
\vspace*{-1.5mm}
Even for the simple case of zero gravity,
this makes the numerical calculation
of large bubbles difficult.
There are, then, two very different scales,
namely the bubble radius ($\overline{\rho}\gg 1$)
and the width of the bubble wall ($\Delta\rho \sim 1$).

Remark, finally,
that the above two characteristics of the $q$-bubble dynamics
are very different from those of Coleman's vacuum
bubble, as mentioned already in Sec.~\ref{sec:Intro}.
Indeed, Coleman's vacuum bubble~\cite{Coleman1977} has no
in-moving disturbance and an essentially constant domain-wall
profile in its rest-frame, energy being supplied by the ``false'' vacuum.

 \newpage
\section{Bubble with gravity}
\label{sec:Bubble-with-gravity}

\subsection{Preliminaries and Ans\"{a}tze}  
\label{subsec:Preliminaries-and-Ansaetze}

From now on, we set  
\beq\label{eq:G-is-GsubN}
G=G_{N}\,,
\eeq
where $G_{N}$ is Newton's gravitational coupling  
constant~\cite{Tanabashi-etal2018}.

The spherically symmetric \textit{Ansatz} for the metric
in Kodama--Schwarzschild
coordinates $(t,\,  r,\,  \theta,\, \phi)$
reads~\cite{AbreuVisser2010}
\beq\label{eq:metric-Ansatz}
g_{\alpha\beta}=  \left[
\text{diag} \left(
-e^{-2\Phi(t,\,r)}\,\left[1 - \frac{2\, G_{N}\, m(t,\,r)}{r}\right],\,
\left[1 - \frac{2\, G_{N}\, m(t,\,r)}{r}\right]^{-1},\,
r^{2},\,r^{2} \,   \sin^{2}\theta   \right)
\right]_{\alpha\beta}\,,  
\eeq
and the spherically symmetric  \textit{Ansatz} for the matter field
is simply
\beq\label{eq:q-field-Ansatz}
q=q(t,\,r)\,.
\eeq

It is a straightforward exercise to insert these
\textit{Ans\"{a}tze} into the
field equations \eqref{eq:nonlin-Klein-Gordon-eq}  
and \eqref{eq:Einstein-eq}
from the action \eqref{eq:action-sigmaLambda-definition-Fq-definition}.
In this way, the reduced nonlinear-Klein--Gordon equation
and the reduced Einstein equations are obtained
(these expressions will be given in Sec.~\ref{subsec:Dimensionless-PDEs}).

\subsection{Dimensionless PDEs}
\label{subsec:Dimensionless-PDEs}

As mentioned in Sec.~\ref{sec:Theory and setup},
specifically in the paragraph above \eqref{eq:epsilon-Ansatz},
we make all variables dimensionless by use of $q_{0}>0$,
which we now take to have the following numerical value:
\beq\label{eq:q0-Planck-scale}
q_{0} 
\equiv g\, (G)^{-1}  
= g\, (G_{N})^{-1} \equiv g\, (E_\text{Planck})^{2}
     \approx g\, \left(1.22 \times 10^{19}\,\text{GeV}
     \right)^{2}\,.  
\eeq
With $q_{0}\equiv (E_\text{$q$-field})^{2}$, the number 
$g$ here can be interpreted as a hierarchy factor,
\beq\label{eq:g-hierarchy-factor}
g=  \left(E_\text{$q$-field}/E_\text{Planck} \right)^{2}\,.
\eeq

Added to our previous dimensionless $q$-field \textit{Ansatz} function
$f(\tau,\,\rho)$,
we now have two dimensionless metric \textit{Ansatz} functions,
making for a total of three \textit{Ansatz} functions:
\beq\label{eq:dimensionless-functions}
\Big\{f(\tau,\,\rho)\,,\Phi(\tau,\,\rho)\,, \mu(\tau,\,\rho)\Big\}\,.
\eeq
A useful definition is
\beq\label{eq:def-B}
B(\tau,\,\rho) \equiv 1-2\,\mu(\tau,\,\rho)/\rho\,,
\eeq
as precisely this combination enters
the metric \textit{Ansatz} \eqref{eq:metric-Ansatz}
by the square bracket factors in $g_{\tau\tau}$ and $g_{\rho\rho}$,
using dimensionless coordinates $\tau$ and $\rho$ instead of $t$ and $r$.

The reduced nonlinear-Klein--Gordon equation corresponds to the
following PDE:
\beq\label{eq:reduced-nonlinear-KG-eq}
\frac{e^{2\Phi}}{B}\,\ddot{f}
-B\,\frac{1}{\rho^{2}}\,\partial_{\rho} \Big(\rho^{2}\,f'\Big)
+\frac{e^{2\Phi}}{B}\, \left(\dot{\Phi}+\frac{2\,\dot{\mu}}{\rho\,B}\right)\,\dot{f}
+\left(
B\,\Phi^\prime
+ \frac{2\,\mu^\prime}{\rho}
- \frac{2\,\mu}{\rho^{2}}
\right)\,f^\prime
=-\frac{d r_{V}}{d f}\,,
\eeq
where an overdot stands for differentiation with respect to
the dimensionless time coordinate
$\tau$ and a prime for differentiation with respect to 
the dimensionless radial coordinate $\rho$.
The reduced Einstein equations give the following first-order PDEs:
\bsubeqs\label{eq:reduced-Einstein-1storder-eqs}
\beqa\label{eq:reduced-Einstein-muprime-eq}
\frac{\mu^\prime}{\rho^{2}}
&=&
4\pi\,g\,  \left[ r_{V} + B\, \frac{1}{2}\,\big(f^\prime\big)^{2}
              +\frac{e^{2\Phi}}{B}
              \,\frac{1}{2}\,\big(\dot{f}\big)^{2}\right] \,,
\\[2mm]
\label{eq:reduced-Einstein-mudot-eq}
\frac{\dot{\mu}}{\rho^{2}\,B} &=&
4\pi\,g\,  f^\prime\,\dot{f}\,,
\\[2mm]
\label{eq:reduced-Einstein-Phiprime-eq}
\frac{\Phi^\prime}{\rho}\,B &=&
8\pi\,g\,  r_{V}- 2\, \mu'/\rho^{2}   \,,
\eeqa
\esubeqs
and the following second-order PDE:
\beq\label{eq:reduced-Einstein-2ndorder-eq}
 \frac{\mu^{\prime\prime}}{\rho}
+\frac{e^{\Phi}}{\rho\,\sqrt{B}}\, \partial_{\rho} \,
 \Big[ \rho\, B^{3/2}\,  e^{-\Phi}\, \Phi^\prime \Big]
+\frac{e^{\Phi}}{\rho}\,\partial_{\tau} \,
 \left[ \frac{e^{\Phi}}{B^{2}}  \,\dot{\mu}\right]=
8\pi\,g\, \left\{ r_{V} + B\, \frac{1}{2}\,\big(f^\prime\big)^{2}
- \frac{e^{2\Phi}}{B}\, \frac{1}{2}\,\big(\dot{f}\big)^{2}
\right\}\,.
\eeq
Note that we have used \eqref{eq:reduced-Einstein-muprime-eq}
to get the expression on the
right-hand side of \eqref{eq:reduced-Einstein-Phiprime-eq}.

The following consistency check holds:
the second-order PDE \eqref{eq:reduced-Einstein-2ndorder-eq}
is solved by the solutions of the
first-order PDEs \eqref{eq:reduced-Einstein-1storder-eqs}
and the second-order PDE \eqref{eq:reduced-nonlinear-KG-eq}.
It is a well-known fact that the same holds for the
reduced ordinary differential equations (ODEs)
of the standard Friedmann--Robertson--Walker universe.
Specifically, the second-order reduced Einstein ODE
 follows from the first-order reduced Einstein ODE
(a.k.a. the Friedmann equation) by use of
the energy-momentum-conservation relations
of the perfect fluid considered.
Ultimately, this redundancy of the reduced field equations
traces back to the invariance of the theory
under general coordinate transformations; cf.
Sec.~15.1, p. 473 of Ref.~\cite{Weinberg1972}.

We can also obtain a useful $g$-independent relation from
\eqref{eq:reduced-Einstein-muprime-eq}
and \eqref{eq:reduced-Einstein-mudot-eq}
in three steps.
First, we extract $\mu^{\prime}$ from \eqref{eq:reduced-Einstein-muprime-eq}
and take the $\tau$ derivative.
Second, we extract $\dot{\mu}$ from \eqref{eq:reduced-Einstein-mudot-eq}
and take the $\rho$ derivative,
Third, we equate the two expressions for $\dot{\mu}^{\prime}$.
The obtained relation is
\beq
\label{eq:relation}
\partial_{\tau} \left(\rho^{2} \left[ r_{V} + B\, \frac{1}{2}\,\big(f^\prime\big)^{2}
              +\frac{e^{2\Phi}}{B}
              \,\frac{1}{2}\,\big(\dot{f}\big)^{2}\right]\right)
=
\partial_{\rho}\,
\left(\rho^{2}\,B\,f^\prime\,\dot{f} \right)\,,
\eeq
which may be interpreted as a current-conservation relation.
Indeed, for the setup of our initial-value problem
(with $f=f_{0}=\text{constant}$ for $\rho \geq \overline{\rho}+1/2$
at $\tau=0$),
the integral of \eqref{eq:relation}
gives the following conserved energy $E$:
\bsubeqs\label{eq:E-def-e-def}
\beqa\label{eq:E-def}
E &=& \sqrt{q_{0}}\, \int_{0}^{\infty} d\rho\;
4\pi \rho^{2} \;e\,,
\\[2mm]
\label{eq:e-def}
e &=& r_{V} + B\, \frac{1}{2}\,\big(f^\prime\big)^{2}
              +\frac{e^{2\Phi}}{B}
              \,\frac{1}{2}\,\big(\dot{f}\big)^{2}\,,
\eeqa
\esubeqs
where the equilibrium value $q_{0}$ of the
$q$ variable in the four-form-field-strength realization \eqref{eq:Fq-definition}
has been used to make lengths and times dimensionless
(Sec.~\ref{sec:Theory and setup}).
Incidentally, the relation \eqref{eq:relation}     
reproduces the reduced nonlinear-Klein--Gordon equation 
\eqref{eq:reduced-nonlinear-KG-eq} 
upon use of \eqref{eq:reduced-Einstein-1storder-eqs}.

Consistent with the expected de-Sitter behavior $m(t,\,r) \propto r^3$  
near the center
and the expected Schwarzschild behavior $m(t,\,r) \sim \text{constant}$
towards spatial infinity,
we take the following boundary conditions on the dimensionless
metric function $\mu(\tau,\,\rho)$:
\bsubeqs\label{eq:nu-Phi-rho-BCS-zero-infty}
\beqa
\mu(\tau,\,0)     &=&0\,,
\\[2mm]
\partial_{\rho}\,\mu(\tau,\,\infty)&=&0\,.
\eeqa
For the other metric function $\Phi(\tau,\,\rho)$, we take
\beqa
\Phi(\tau,\,0)&=&0\,,
\\[2mm]
\partial_{\rho}\,\Phi(\tau,\,\infty)&=&0\,.
\eeqa
\esubeqs
The boundary conditions on $f(\tau,\,\rho)$ have already been
given in \eqref{eq:f-bcs-rho-origin}
and \eqref{eq:f-bcs-rho-infinity}.
From the boundary conditions \eqref{eq:nu-Phi-rho-BCS-zero-infty},
we find that the reduced Einstein
equations \eqref{eq:reduced-Einstein-1storder-eqs}
and \eqref{eq:reduced-Einstein-2ndorder-eq},
for the case $g=0$, give
$\mu(\tau,\,\rho)=0$ and $\Phi(\tau,\,\rho)=0$,
so that \eqref{eq:reduced-nonlinear-KG-eq}
reproduces the flat-spacetime PDE
\eqref{eq:flat-spacetime-reduced-nonlinear-KG-eq}.

\subsection{Numerics}
\label{subsec:Numerics}

\subsubsection{Numerical procedure}
\label{subsubsec:Numerical-procedure}

Finding the numerical solution of the
PDEs \eqref{eq:reduced-nonlinear-KG-eq},
\eqref{eq:reduced-Einstein-1storder-eqs},
and \eqref{eq:reduced-Einstein-2ndorder-eq}
is a nontrivial task. In the following local approach,
we are inspired by the discussion 
of App.~\ref{app:Integro-differential-eqs}.

The coordinates $\rho$ and $\tau$  are put on a finite grid
with $N_{\rho}$ and $N_{\tau}=2\,N_{\rho}$ points, respectively.
The PDEs \eqref{eq:reduced-nonlinear-KG-eq},
\eqref{eq:reduced-Einstein-Phiprime-eq},
and \eqref{eq:reduced-Einstein-2ndorder-eq}   
are then solved with time-derivatives of $f$ and $\mu$
replaced by forward time-differences
and the time-derivative of $\Phi$
replaced by a backward time-difference.

\subsubsection{Numerical solutions}
\label{subsubsec:Numerical-solutions}

For the presentation of our numerical results, we will employ
time-slice plots
(cf. Fig.~\ref{fig:two-panel-f-rV-timeslices-bubble-for-g-0})
rather than surface plots
(cf. Figs.~\ref{fig:two-panel-f-rV-surfaceplots-start-bubble-for-g-0} 
and \ref{fig:two-panel-f-rV-surfaceplots-bubble-for-g-0}).
The various time-slices will be collected in a single
plot by color-coding the different time values.

\begin{figure}[p]
\vspace*{-2mm}
\begin{center}   
\includegraphics[width=0.689\textwidth]{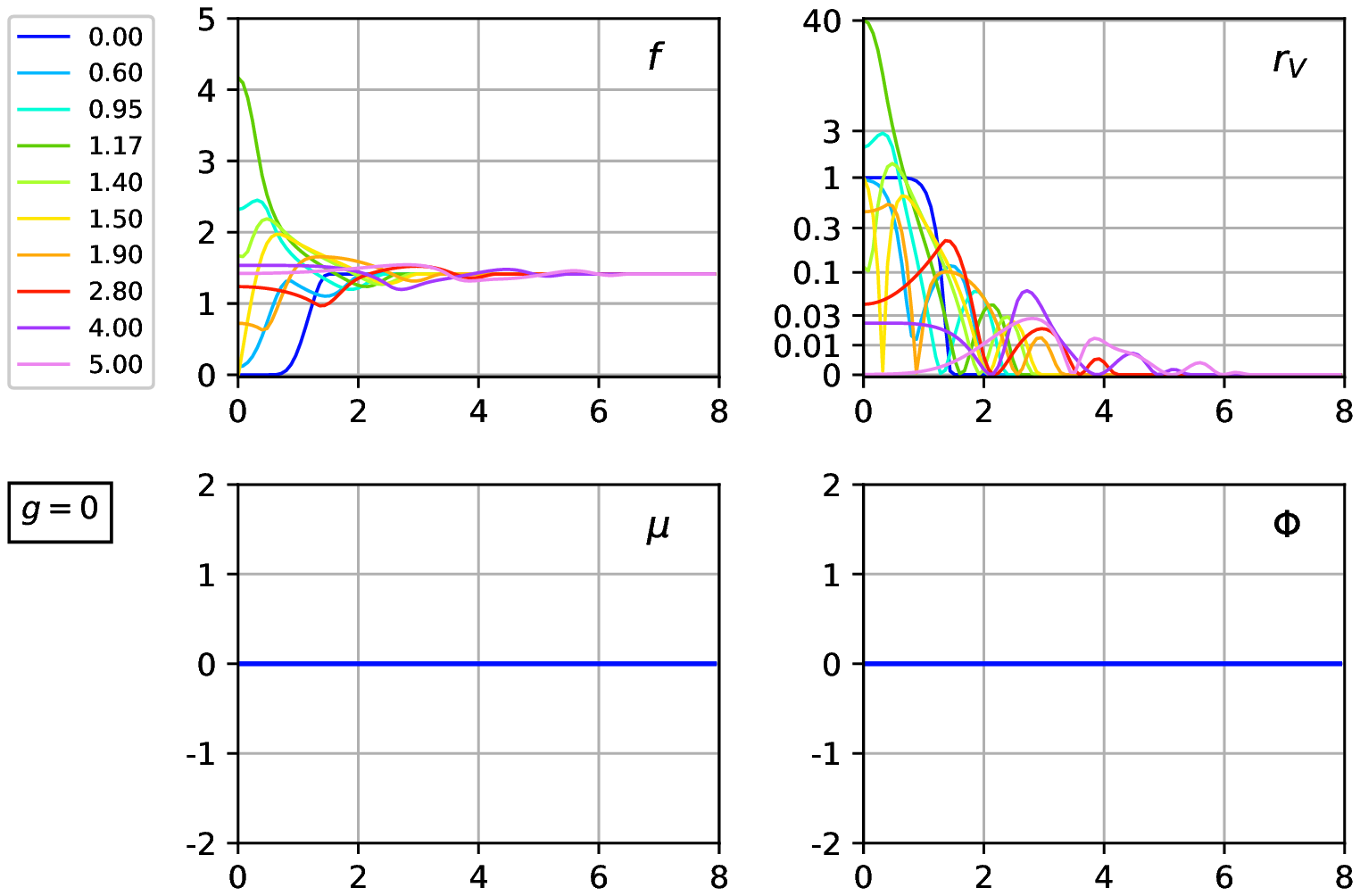}
\end{center}
\vspace*{-5mm}
\caption{Numerical solution of the PDEs \eqref{eq:reduced-nonlinear-KG-eq},
\eqref{eq:reduced-Einstein-1storder-eqs}, and \eqref{eq:reduced-Einstein-2ndorder-eq}:
plots of $f(\tau,\,\rho)$, $r_{V}(\tau,\,\rho)$,
$\mu(\tau,\,\rho)$, and $\Phi(\tau,\,\rho)$
at different time slices,
with $\tau$ values given in the legend on the left-hand side.
The model parameters  are $\lambda=1$
and $g=0$. The initial values are: $f(0,\rho)$ as given by
\eqref{eq:f-start} with $\overline{\rho}=1$
and $\dot{f}(0,\rho)=0$.
The metric functions $\mu(\tau,\,\rho)$ and $\Phi(\tau,\,\rho)$
vanish identically.
The vacuum energy density $r_{V}$ is plotted as $\log_{10}(r_{V}+0.01)$.
}
\label{fig:four-panel-f-rV-mu-Phi-for-g-0}
\vspace*{25mm}
%
\vspace*{-0cm}
\begin{center}   
\includegraphics[width=0.689\textwidth]{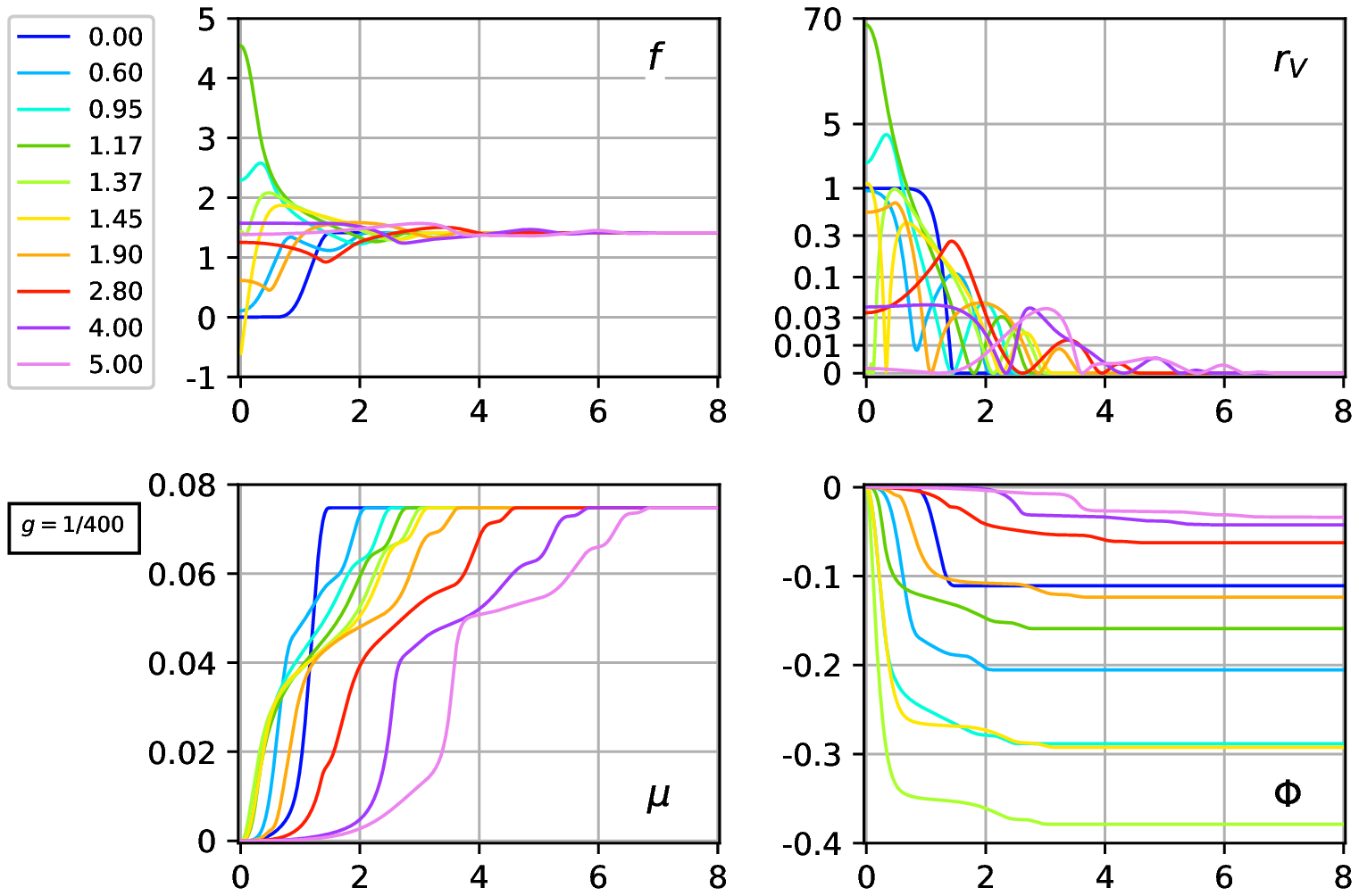}
\end{center}
\vspace*{-5mm}
\caption{Same as Fig.~\ref{fig:four-panel-f-rV-mu-Phi-for-g-0},
again with $\lambda=1$ but now for $g=0.0025$.   
The initial values are: $f(0,\rho)$ as given by
\eqref{eq:f-start} with $\overline{\rho}=1$,
$\dot{f}(0,\rho)=0$, $\mu(0,\rho)$ from
\eqref{eq:reduced-Einstein-muprime-eq},
and $\Phi(0,\rho)$
from \eqref{eq:reduced-Einstein-Phiprime-eq}.
}
\label{fig:four-panel-f-rV-mu-Phi-for-g-0pt0025}
\vspace*{0cm}
\end{figure}

The numerical solution for $g=0$ (Fig.~\ref{fig:four-panel-f-rV-mu-Phi-for-g-0})
can now be compared with the one
for $g=1/400$ (Fig.~\ref{fig:four-panel-f-rV-mu-Phi-for-g-0pt0025}).
For the last case, in particular,
it has been verified that the numerically obtained functions
$f(\tau,\,\rho)$, $\mu(\tau,\,\rho)$, and $\Phi(\tau,\,\rho)$
give residuals of the 
first-order PDEs \eqref{eq:reduced-Einstein-muprime-eq}  
and \eqref{eq:reduced-Einstein-mudot-eq}
that drop to zero as the number of grid points increases.

\begin{figure}[p]
\vspace*{-2mm}
\begin{center}   
\includegraphics[width=0.689\textwidth]{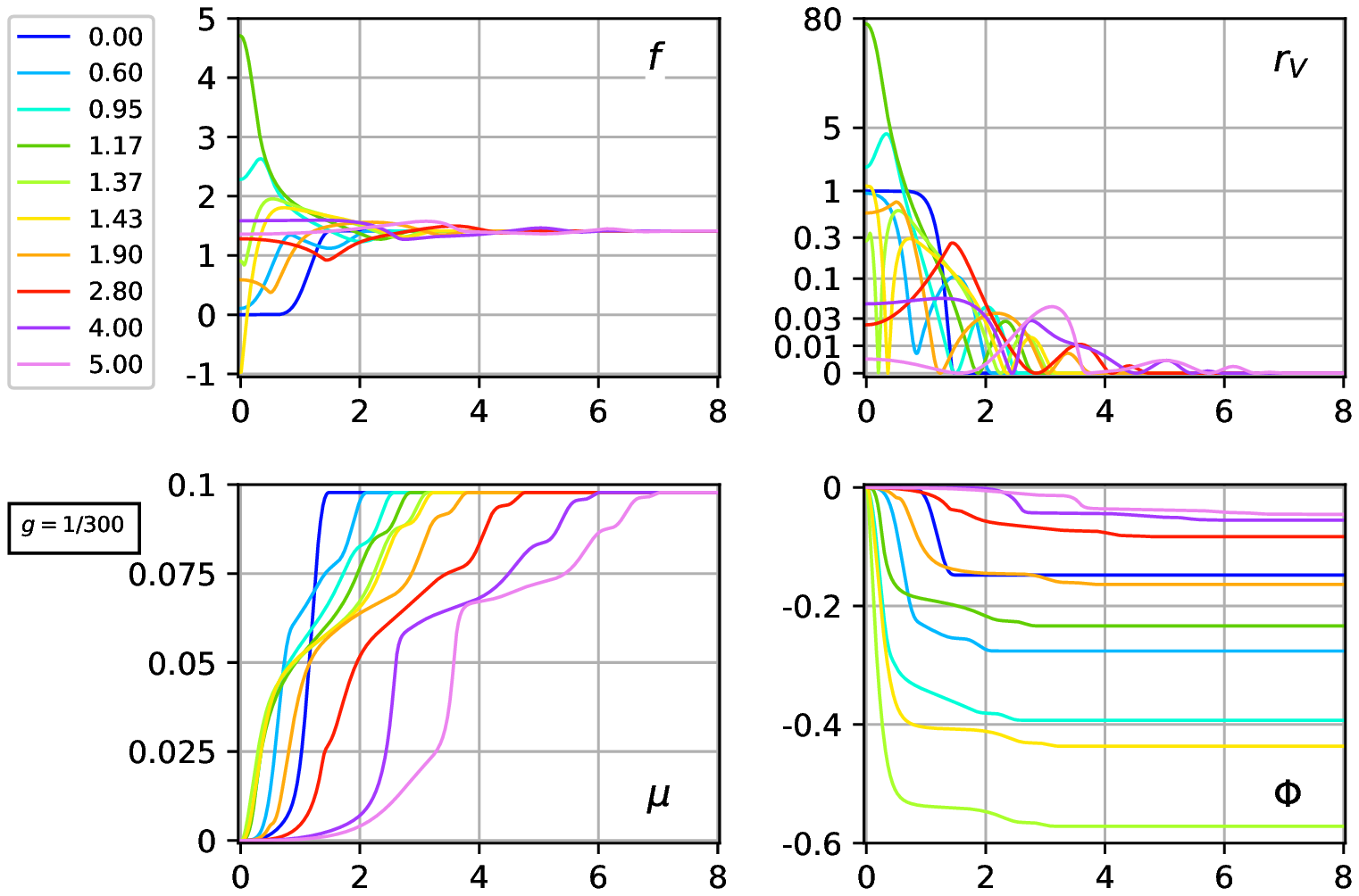}
\end{center}
\vspace*{-5mm}
\caption{Same as Fig.~\ref{fig:four-panel-f-rV-mu-Phi-for-g-0pt0025},
but now for $g=0.0033\!\!\not{\!3}$. 
}
\label{fig:four-panel-f-rV-mu-Phi-for-g-0pt0033}
\vspace*{6mm}
\begin{center}
\includegraphics[width=1\textwidth]{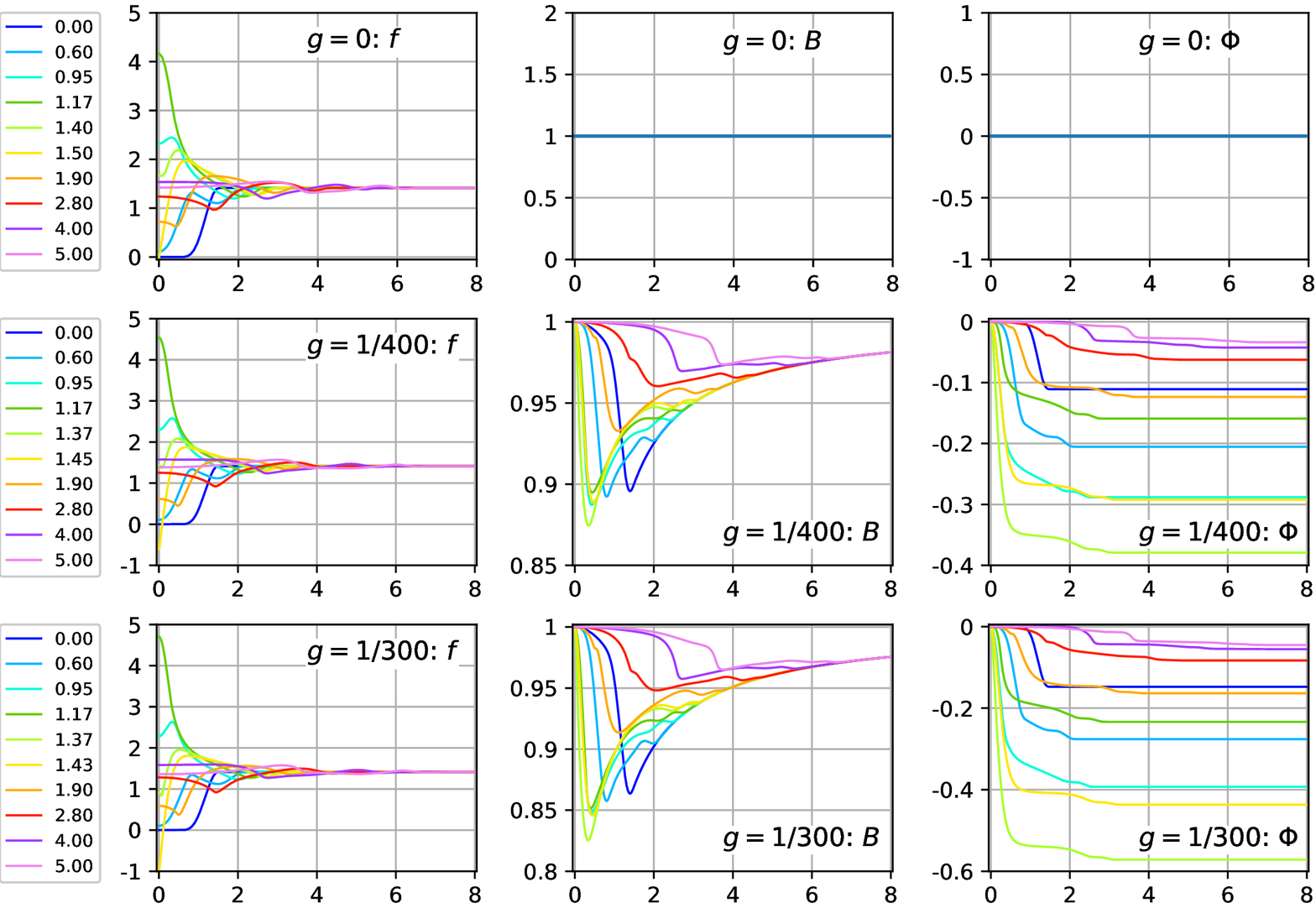}
\end{center}
\vspace*{-5mm}
\caption{Comparison of the numerical solutions 
from Figs.~\ref{fig:four-panel-f-rV-mu-Phi-for-g-0}--\ref{fig:four-panel-f-rV-mu-Phi-for-g-0pt0033},
showing, in particular, the behavior
of the quantity $B$ from \eqref{eq:def-B},
which enters the metric \eqref{eq:metric-Ansatz}.
The various time-slices at the three different values of $g$
are given by the respective legends on the left.}
\label{fig:nine-panel-f-B-Phi-for-g-0-0pt025-0pt0033}
\vspace*{0cm}
\end{figure}

\begin{figure}[t]
\begin{center}
\includegraphics[width=0.689\textwidth]{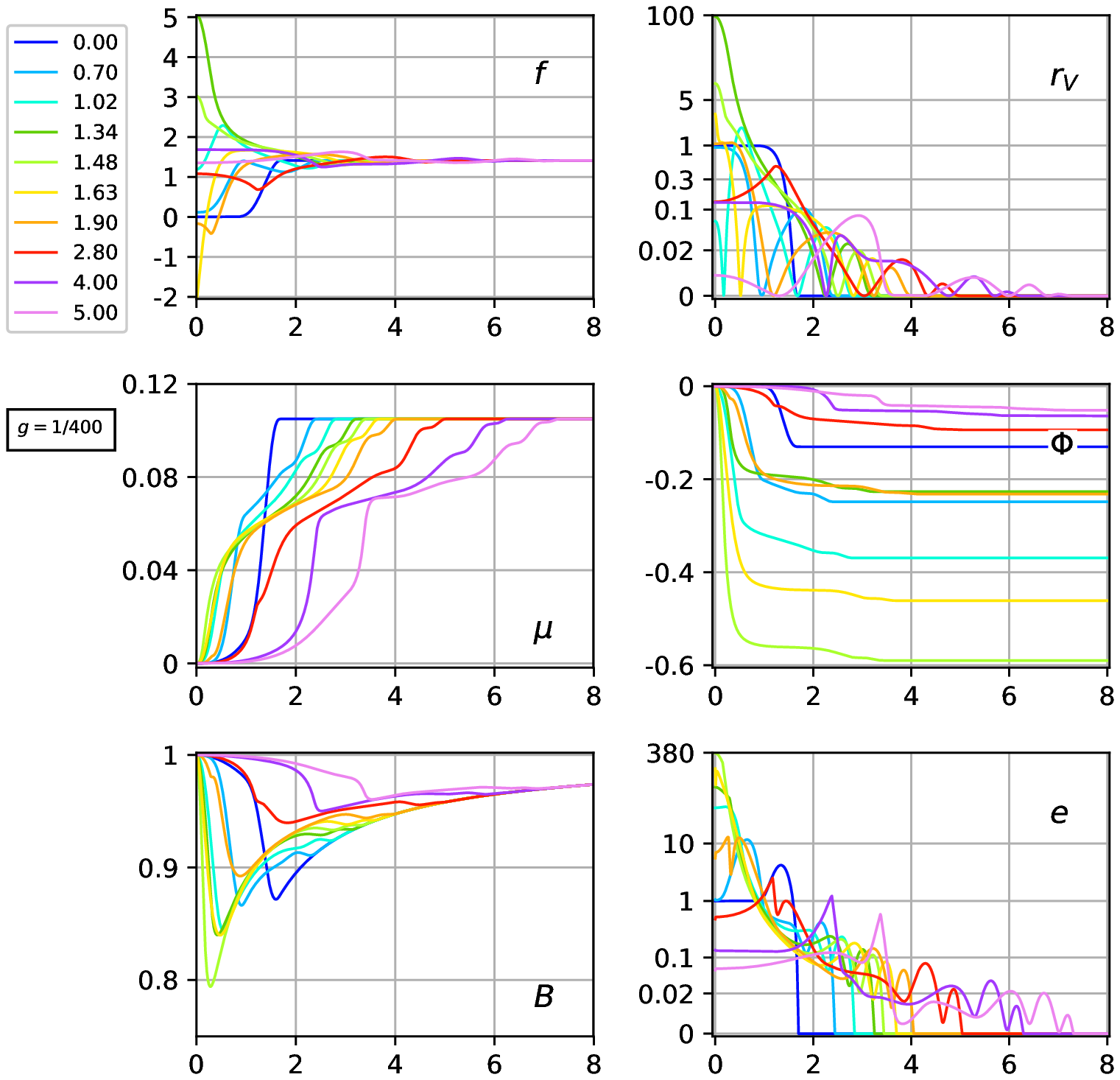}
\end{center}
\vspace*{-5mm}
\caption{Same as Fig.~\ref{fig:four-panel-f-rV-mu-Phi-for-g-0pt0025},
but now with $\overline{\rho} = 1.2$ instead of $\overline{\rho} = 1$.
Two additional panels show the metric quantity $B(\tau,\,\rho)$ from \eqref{eq:def-B}
and the energy density $e(\tau,\,\rho)$ from \eqref{eq:e-def}.
The energy densities $e$ and $r_{V}$ are plotted using the same   
scale function $\log_{10}(x+0.005)$  
but with different over-all factors.
The $r_{V}$ peak at $(\tau,\,\rho)\sim (1.34,\,0)$
of the $\overline{\rho} = 1.2$ solution
is significantly larger
than the corresponding peak of the $\overline{\rho} = 1$ solution in
Fig.~\ref{fig:four-panel-f-rV-mu-Phi-for-g-0pt0025}.
Similarly, the $B$ dip of the $\overline{\rho} = 1.2$ solution
is significantly lower
than the corresponding dip of the $\overline{\rho} = 1$ solution
in the $B$ panel of the middle row of
Fig.~\ref{fig:nine-panel-f-B-Phi-for-g-0-0pt025-0pt0033}.
}
\label{fig:six-panel-f-rV-mu-Phi-B-e-for-g-0pt0025}
\vspace*{0cm}
\end{figure}

For somewhat larger $g$ values,
a Schwarzschild-type horizon is formed,
as the energy density $e$ becomes large close to the center $\rho=0$.
This horizon is
apparently different from a de-Sitter-type horizon which
arises from a constant vacuum energy density far away from the
center; see App.~\ref{app:Bubble-interior} for a brief
discussion of the de-Sitter-type spacetime near the $q$-bubble origin.
With the setup and boundary conditions from
Fig.~\ref{fig:four-panel-f-rV-mu-Phi-for-g-0pt0025},
we estimate horizon formation to occur for $g \gtrsim 0.006$.  
The regular numerical solution at $g=1/300$   
is shown in Fig.~\ref{fig:four-panel-f-rV-mu-Phi-for-g-0pt0033}.
The evolution towards the formation of a horizon,
with $B(\tau,\,\rho)$ from \eqref{eq:def-B} dipping to zero,
is illustrated in Fig.~\ref{fig:nine-panel-f-B-Phi-for-g-0-0pt025-0pt0033}.

For a large bubble, we expect that, from the ingoing $r_{V}$ disturbance 
(cf. Figs.~\ref{fig:two-panel-f-rV-surfaceplots-start-bubble-for-g-0}
and \ref{fig:two-panel-f-rV-surfaceplots-bubble-for-g-0}), 
the $r_{V}$ peak close to the
origin will be higher than the one for a small bubble.
This behavior is confirmed by
comparing Fig.~\ref{fig:six-panel-f-rV-mu-Phi-B-e-for-g-0pt0025}
with Fig.~\ref{fig:four-panel-f-rV-mu-Phi-for-g-0pt0025}.
The $f$-panel in Fig.~\ref{fig:six-panel-f-rV-mu-Phi-B-e-for-g-0pt0025}
also shows that the quantities $(\dot{f})^2$ and $(f')^2$ are large
at $(\tau,\,\rho) \sim (1.5,\,0)$,   
with both terms contributing
significantly to the energy density $e$ close to the center $\rho=0$.

\subsection{Discussion}
\label{subsec:Discussion-with-grav}

The numerical results of Sec.~\ref{subsubsec:Numerical-solutions}
show that the vacuum energy density of
a nonequilibrium $q$-bubble
embedded in the equilibrium vacuum with $q=q_{0}=\text{constant}$
evolves in a complicated way. For a sufficiently small
$q$-bubble,
part of the vacuum energy density $r_{V}$ of the bubble wall,
first, moves inwards towards the center and, then, rapidly disperses
(cf. Figs.~\ref{fig:two-panel-f-rV-surfaceplots-start-bubble-for-g-0}
and \ref{fig:two-panel-f-rV-surfaceplots-bubble-for-g-0}).

The numerical calculations were performed
for the case with gravity turned off ($G=0$) and turned on ($G > 0$).   
Qualitatively, the main effect of gravity
is to give a larger maximum value of the vacuum energy density
at the center $\rho=0$ (compare the $r_{V}$ panels of
Fig.~\ref{fig:four-panel-f-rV-mu-Phi-for-g-0}
and \ref{fig:four-panel-f-rV-mu-Phi-for-g-0pt0025}).

If the hierarchy ratio $g$ from \eqref{eq:g-hierarchy-factor}
is approximately equal to or somewhat above $0.006$,   
the particular solution develops a Schwarzschild-type horizon
near the center $\rho=0$ and different
coordinates need to be chosen (cf. App.~\ref{app:Bubble-interior}).
We postpone this analysis to a future publication, as the focus
of the present article is on the dispersion of vacuum energy if
the Big Bang occurs in a finite region of space surrounded by
equilibrium vacuum (where any form of initial vacuum energy has already been 
cancelled~\cite{KV2008a,KV2008b,KV2016-Lambda-cancellation}).

\section{Conclusions}   
\label{sec:Conclusions}

In the present article, we have
obtained a first glimpse of the inhomogeneous dynamics of the
gravitating vacuum energy density $\rho_{V}(q)$
as described by the vacuum variable $q$ originating from
a four-form field strength
(earlier work~\cite{KV2008b,KV2016-Lambda-cancellation} 
considered the time-evolution     
of spatially-constant $q$-fields). 
In this new probe of $q$-theory,
we start from a large vacuum energy density in
a finite region of space surrounded by equilibrium vacuum, and
follow the time evolution of the vacuum energy density.

Our numerical results show the
possibility of obtaining different evolution scenarios
depending on the initial
conditions and the parameters of the vacuum energy.
These results suggest that there may be
de-Sitter expansion within a finite region of space,
gravitational collapse of the vacuum medium
with the formation of a singularity, and
formation of cosmological and/or black-hole horizons.

It may also be of interest to study the vacuum structure at the black hole 
singularity. The singularity may be smoothened, as the gravitational 
coupling depends, in general, on the value of the $q$ variable and 
gravity may be effectively turned off near the center. 
We leave this study to a future investigation.

\begin{acknowledgments}
The work of G.E.V. has been supported by the European Research Council
(ERC) under the European Union's Horizon 2020 research and innovation
programme (Grant Agreement No. 694248).
O.P.S is supported by the Beca Externa Jovenes Investigadores
of CONICET.
\end{acknowledgments}

\begin{appendix}
\section{Integro-differential equations}
\label{app:Integro-differential-eqs}

The role of $\Phi(\tau,\,\rho)$ in the PDEs
\eqref{eq:reduced-nonlinear-KG-eq},
\eqref{eq:reduced-Einstein-1storder-eqs}, and
\eqref{eq:reduced-Einstein-2ndorder-eq} is rather subtle.
From \eqref{eq:reduced-Einstein-Phiprime-eq}, we have
\beq\label{eq:Phiprime-eq}
\Phi' = \frac{8\pi\,g\,  \rho\, r_{V}- 2\, \mu'/\rho}{1-2\,\mu/\rho}\,,
\eeq
which can be integrated to give 
\beq\label{eq:Phi-integral}
\widehat{\Phi}(\tau,\,\rho) = \int_{0}^{\rho}\,d\widetilde{\rho}\;
\frac{8\pi\,g\, \widetilde{\rho}\, r_{V}[f(\tau,\,\widetilde{\rho})]
     - 2\, \mu'(\tau,\,\widetilde{\rho})/\widetilde{\rho}}
     {1-2\,\mu(\tau,\,\widetilde{\rho})/\widetilde{\rho}}\,,
\eeq
where the prime in the numerator of the integrand stands for
differentiation with respect to $\widetilde{\rho}$.
Hence, $\widehat{\Phi}(\tau,\,\rho)$
is determined \emph{nonlocally} by the functions
$f(\tau,\,\widetilde{\rho})$ and $\mu(\tau,\,\widetilde{\rho})$
at the same time slice $\tau$.

The PDEs \eqref{eq:reduced-nonlinear-KG-eq}
and \eqref{eq:reduced-Einstein-2ndorder-eq}
still involve $\Phi$  and its time-derivative $\dot{\Phi}$
[the spatial derivatives
$\Phi'$ and $\Phi''$ can be eliminated by use of
\eqref{eq:reduced-Einstein-Phiprime-eq}].
Replacing $\Phi(\tau,\,\rho)$ by $\widehat{\Phi}(\tau,\,\rho)$
from \eqref{eq:Phi-integral}, these two equations become
\emph{integro-differential equations} solely involving the
functions $f(\tau,\,\rho)$ and $\mu(\tau,\,\rho)$.
Explicitly, these equations read:
\bsubeqs\label{eq:integro-diff-eqs}
\beqa\label{eq:nonlinear-KG-integro-diff-eq}
\hspace*{-15mm}
&&
\frac{e^{2\widehat{\Phi}}}{B}\,\ddot{f}
-B\,\frac{1}{\rho^{2}}\,\partial_{\rho} \Big(\rho^{2}\,f'\Big)
+\frac{e^{2\widehat{\Phi}}}{B}\,
\left(\partial_{\tau}\widehat{\Phi}  
+\frac{2\,\dot{\mu}}{\rho\,B}\right)\,\dot{f}
+\left(
B\,\widehat{\Phi}^\prime
+ \frac{2\,\mu^\prime}{\rho}
- \frac{2\,\mu}{\rho^{2}}
\right)\,f^\prime
=-\frac{d r_{V}}{d f}\,,
\\[2mm]
\label{eq:Einstein-integro-diff-eq}
\hspace*{-15mm}
&&
 \frac{\mu^{\prime\prime}}{\rho}
+\frac{e^{\widehat{\Phi}}}{\rho\,\sqrt{B}}\, \partial_{\rho}
 \Big[ \rho\, B^{3/2}\,  e^{-\widehat{\Phi}}\, \widehat{\Phi}^\prime \Big]
+\frac{e^{\widehat{\Phi}}}{\rho}\partial_{\tau}
 \left[ \frac{e^{\widehat{\Phi}}}{B^{2}}  \,\dot{\mu}\right]=
8\pi\,g \left\{ r_{V} + B\, \frac{1}{2}\,\big(f^\prime\big)^{2}
- \frac{e^{2\widehat{\Phi}}}{B}\, \frac{1}{2}\,\big(\dot{f}\big)^{2}
\right\},
\eeqa
\esubeqs
with $\widehat{\Phi}$ given by the expression \eqref{eq:Phi-integral}
and $\partial_{\tau}\widehat{\Phi}$
having the $\tau$-derivative pulled inside the $\widetilde{\rho}$ integral.

\section{Bubble interior}
\label{app:Bubble-interior}
\setcounter{section}{1}  
\setcounter{equation}{3}

The $q$-bubble setup considered in this article has
a start configuration $f(0,\,\rho)$ determined by
\eqref{eq:f-start} and the further initial condition
$\dot{f}(0,\,\rho)=0$. Then, the reduced
field equation \eqref{eq:reduced-Einstein-muprime-eq}
gives that the metric \textit{Ansatz}
function $\mu(\tau,\,\rho)$ behaves near the center
as $\mu(\tau,\,\rho)\propto \rho^3$.
This behavior of $\mu$
allows for the following definition of the quantity $h(\tau)$:
\beqa\label{eq:h-definition-near-bubble-origin}
\lim_{\rho \to 0}
\frac{2\, \mu(\tau,\,\rho)}{\rho^3}
&\equiv&
h^{2}(\tau).
\eeqa
Near the spacetime origin  ($\rho=\tau=0$) of the $q$-bubble considered,
we have
\bsubeqs\label{eq:h-Phi-near-bubble-origin}
\beqa
h^{2}(\tau)   &\sim& h^{2}(0) \equiv h_{0}^{2}\,,
\\[2mm]
\Phi(0,\,0)  &\sim& 0\,,
\\[2mm]
r_{V}(0,\,0) &\sim& r_{V0}  >0 \,,
\eeqa
\esubeqs
with constants $r_{V0}$ and $h_{0}$. In fact, the reduced
field equation \eqref{eq:reduced-Einstein-muprime-eq} gives
\beq\label{eq:h0square-results}
h_{0}^{2}= (8\pi/3) \,g\,r_{V0}\,,
\eeq
where $g$ has been defined in \eqref{eq:q0-Planck-scale}.
The resemblance of \eqref{eq:h0square-results}
with the spatially flat Friedmann equation~\cite{Weinberg1972} 
of a universe with constant vacuum energy will become clear later on.

Writing the metric \eqref{eq:metric-Ansatz}
in terms of dimensionless variables gives
\beqa\label{eq:dimensionless-metric}
ds^{2} &=&
-e^{-2\Phi(\tau,\,\rho)}\,\left[1 - \frac{2\, \mu(\tau,\,\rho)}{\rho}\right]\,d\tau^{2}
+\left[1 - \frac{2\, \mu(\tau,\,\rho)}{\rho}\right]^{-1}\,d\rho^{2}
\nonumber\\[1mm]
&&
+\rho^{2} \Big(d\theta^{2} + \sin^{2}\theta\,d\phi^{2} \Big)\,.
\eeqa
With the behavior \eqref{eq:h-definition-near-bubble-origin}
and \eqref{eq:h-Phi-near-bubble-origin},
the metric \eqref{eq:dimensionless-metric} 
near the spacetime origin of the $q$-bubble ($\rho=\tau=0$) becomes
\beqa\label{eq:dimensionless-metric-near-bubble-origin}
ds^{2}\,\Big|_\text{origin} &\sim&
-\Big[1 - h_{0}^{2}\,\rho^{2}\Big]\,d\tau^{2}
+\Big[1 - h_{0}^{2}\,\rho^{2}\Big]^{-1}\,d\rho^{2}
+\rho^{2} \Big(d\theta^{2} + \sin^{2}\theta\,d\phi^{2} \Big)\,,
\eeqa
which corresponds to the metric of de-Sitter spacetime
in so-called static
coordinates~\cite{Schroedinger1956,HawkingEllis1973,BirrellDavies1982}.
Note that, if $\rho$ were allowed to be large enough,
the metric on the right-hand side
of \eqref{eq:dimensionless-metric-near-bubble-origin} would
display a coordinate singularity at $\rho=1/h_{0}$.

Now, introduce new dimensionless coordinates (denoted by a hat)
from the following relations:
\bsubeqs\label{eq:hat-coordinates}
\beqa
\exp(h_{0}\,\widehat{\tau})
&=&
\sqrt{(1- h_{0}^2\,\rho^2)}\,
\Big[ \cosh(h_{0}\,\tau) + \sinh(h_{0}\,\tau) \Big]\,,
\\[2mm]
h_{0}\,\widehat{z} &=& \frac{h_{0}\,\rho\,\cos\theta}{\sqrt{(1- h_{0}^2\,\rho^2)}\,
\Big[ \cosh(h_{0}\,\tau) + \sinh(h_{0}\,\tau) \Big]}\,,
\\[2mm]
h_{0}\,\widehat{y} &=& \frac{h_{0}\,\rho\,\sin\theta\,\cos{\phi}}{\sqrt{(1- h_{0}^2\,\rho^2)}\,
\Big[ \cosh(h_{0}\,\tau) + \sinh(h_{0}\,\tau) \Big]}\,,
\\[2mm]
h_{0}\,\widehat{x} &=& \frac{h_{0}\,\rho\,\sin\theta\,\sin{\phi}}{\sqrt{(1- h_{0}^2\,\rho^2)}\,
\Big[ \cosh(h_{0}\,\tau) + \sinh(h_{0}\,\tau) \Big]}\,.
\eeqa
\esubeqs
With these new coordinates, the metric
\eqref{eq:dimensionless-metric-near-bubble-origin}
near the spacetime origin of the $q$-bubble
($\widehat{\tau}=\widehat{z}=\widehat{y}=\widehat{x}=0$)
becomes
\bsubeqs\label{eq:dimensionless-metric-near-bubble-origin-hat-coordinates}
\beqa\label{eq:FRW-metric}
ds^{2}\Big|_\text{origin}
&\sim&
-(d\widehat{\tau})^{2}
+ \big[a(\widehat{\tau})\big]^{2}\,
\Big[(d\widehat{x})^{2} + (d\widehat{y})^{2} +(d\widehat{z})^{2} \Big]\,,
\\[2mm]
\label{eq:a-FRW-def}
a(\widehat{\tau}) &\equiv& \exp(h_{0}\,\widehat{\tau})\,.
\eeqa
\esubeqs
Note that the spatially-flat Robertson--Walker metric
on the right-hand side of \eqref{eq:FRW-metric}
with the scale factor \eqref{eq:a-FRW-def}
no longer has the nontrivial coordinate singularity.
From the scale factor $a(\widehat{\tau})$ in
\eqref{eq:a-FRW-def}, we obtain $(da/d\widehat{\tau})/a = h_{0}$,
so that the quantity $h_{0}$, which was originally defined
by \eqref{eq:h-definition-near-bubble-origin}
and \eqref{eq:h-Phi-near-bubble-origin},
can be interpreted as a Hubble constant.
The scale factor $a(\widehat{\tau})$
of \eqref{eq:dimensionless-metric-near-bubble-origin-hat-coordinates}
displays, for $h_{0}\,\widehat{\tau} \gg 1$, the well-known
exponential expansion of de-Sitter
spacetime~\cite{Schroedinger1956,HawkingEllis1973,BirrellDavies1982}.

The numerical solution of
Fig.~\ref{fig:four-panel-f-rV-mu-Phi-for-g-0pt0025}, however,
has $h_{0} \approx 0.14$ 
for $\tau \lesssim 0.4$   
and does not show the exponential expansion.
Needed is an initial bubble \eqref{eq:f-start} with $\overline{\rho}\gg 1$
(the required order of magnitude for $\overline{\rho}$
is $1/h_{0} \sim 1/\sqrt{g\,r_{V0}}\,$).
But there are three problems with such large bubbles.
First, as noted in Sec.~\ref{subsec:Discussion-without-grav},
the numerics of a large $q$-bubble is challenging.

Second, the coordinate singularity
of \eqref{eq:dimensionless-metric-near-bubble-origin}
at $\rho=1/h_{0}$ suggests that the metric \textit{Ansatz}
\eqref{eq:dimensionless-metric} is inappropriate.
Most likely, this problem can
be evaded by use of another metric \textit{Ansatz},
possibly inspired by Painlev\'{e}--Gullstrand
coordinates~\cite{Painleve1921,Gullstrand1922,MartelPoisson2000,%
Volovik2009}.

Third, large bubbles may give gravitational collapse
close to the center $\rho=0$, as the energy density $e$ from \eqref{eq:e-def}
becomes large at the center.
See Fig.~\ref{fig:two-panel-f-rV-surfaceplots-start-bubble-for-g-0},
where the initial ($\tau\sim 0$) bubble-wall disturbance 
of the vacuum energy density $r_{V}$
separates around $\tau\sim 0.3$ into an outgoing and ingoing disturbance,
the latter giving a peak of $r_{V}$ at $\rho=0$ for $\tau\sim 1$.
See also Fig.~\ref{fig:six-panel-f-rV-mu-Phi-B-e-for-g-0pt0025},
which shows that the numerical solution with
a somewhat larger value of $\overline{\rho}$
has a significantly larger $r_{V}$ peak at the origin
than the  numerical solution of
Fig.~\ref{fig:four-panel-f-rV-mu-Phi-for-g-0pt0025}.

\end{appendix}


\end{document}